\documentclass{IEEEtran4PSCC}
\ifCLASSINFOpdf
\else
\fi
\usepackage[cmex10]{amsmath}

\usepackage[utf8]{inputenc}

\usepackage{tikz,forest}
\usepackage{pbox}
\usetikzlibrary{arrows,decorations.pathmorphing,backgrounds,fit,positioning,shapes.symbols,chains, shapes,decorations}
\usepackage{mathtools}
\usepackage{cite}
\usepackage{float}
\usepackage{multicol}
\usepackage{multirow}
\usepackage{hyperref}
\usepackage{graphicx}
\usepackage{amsfonts}
\usepackage{amsmath}
\usepackage{color}
\usepackage{epstopdf}
\usepackage{caption}
\usepackage{subcaption}
\usepackage{mathtools, nccmath}
\usepackage{multirow}
\usepackage{bbm}
\usepackage{marginnote}
\usepackage{tikz}
\usepackage{textcomp}
\usepackage{multirow}
\usetikzlibrary{arrows,decorations.pathmorphing,backgrounds,fit,positioning,shapes.symbols,chains, shapes,snakes}
\usepackage{amsthm}

\usepackage{comment}

\begin{document}
%
\title{\huge Chance-Constrained ADMM  Approach for Decentralized Control of Distributed Energy Resources}

\vspace{-0.4cm}
\author{
\IEEEauthorblockN{Ali Hassan \IEEEauthorrefmark{1}, Yury Dvorkin \IEEEauthorrefmark{1}, Deepjyoti Deka \IEEEauthorrefmark{2} and
Michael Chertkov \IEEEauthorrefmark{2}\IEEEauthorrefmark{3}}
\IEEEauthorblockA{\IEEEauthorrefmark{1} Department of Electrical and Computer Engineering,
Tandon School of Engineering,\\
New York University,
New York, NY, USA,
(ah3909,dvorkin)@nyu.edu }
\IEEEauthorblockA{\IEEEauthorrefmark{2} Theory Division,
Los Alamos National Laboratory,\\
Los Alamos, NM, USA,
(deepjyoti, chertkov)@lanl.gov }
\IEEEauthorblockA{\IEEEauthorrefmark{3} Skolkovo Institute of Science and Technology, 143026 Moscow, Russia }
}
\vspace{-0.4cm}


\maketitle


\begin{abstract}

Distribution systems are undergoing a dramatic transition from a passive circuit that routinely  disseminates electric power among downstream nodes to the system with distributed energy resources. The distributed energy resources come in a variety of technologies and typically include photovoltaic (PV) arrays, thermostatically controlled loads, energy storage units. Often these resources are interfaced with the system via inverters that can adjust active and reactive power injections, thus supporting the  operational performance of the system. This paper designs a control policy  for such inverters using the local power flow measurements. The control actuates active and reactive power injections of the inverter-based distributed energy resources. This strategy is then incorporated into a chance-constrained, decentralized optimal power flow formulation to maintain voltage levels and power flows within their limits and to mitigate the volatility of (PV) resources. \textcolor{black}{Our method is shown to improve voltage regulation compliance and reduce power losses.}

\end{abstract}


\thanksto{The work of AH and YD was supported in part by the New York State Energy Research and Development Agency under the project ``DER-centric operating and planning tools for power systems''. The work of DD and MC was supported by funding from the U.S. DOE/OE as part of the DOE Grid Modernization Initiative.}

\section{Introduction}
The continuing proliferation of distributed energy resources (e.g., PV arrays, thermostatically controlled loads, energy storage units)  is transforming the nature of distribution systems. Instead of the passive role of a  mediator between high-voltage transmission systems and low-voltage electricity end-users, distribution systems are in a unique position to leverage distributed energy resources to reduce their operating cost and unlock new revenue streams by  providing service to the transmission system and enabling fine-grain power delivery,  \cite{doe_reprot_2012, bragin}.  The success of this transformation  depends on a coinfluence of such factors as the ability to efficiently deal with high variability of photovoltaic (PV) resources and to maintain power flows and voltage magnitudes in distribution systems within an acceptable range. 
This paper studies if enhanced control policies on inverter-tied distribution energy resources (e.g., PV arrays, thermostatically controlled loads, energy storage units) can provide a sufficient degree of dispatch flexibility to comply with power flow and voltage limits in distribution systems.

Distributed energy resources are technically capable  of providing services to both distribution and transmission systems, see  \cite{Joos_react}. However, there are a few noticeable differences. First, the provision of grid support services in the transmission system is typically organized in a centralized manner and requires that eligible resources meet a certain nameplate capacity threshold. This requirement disqualifies individual grid-scale distributed energy resources, let alone small-scale resources such as residential PV arrays. On the \textcolor{black}{other} hand, distribution system are operated in a less hierarchical manner, which allows for small-scale resources to engage in the provision of grid support services. To enable the provision of these services on a massive scale, i.e. with high penetrations of distributed energy resources, there is a pressing need to devise decentralized control policies. The decentralized framework assumes that distributed energy resources can react to changing operating conditions in the distribution system based on local measurements and without the intervention of the centralized operator. We refer interested readers to \cite{volt_review} for a comprehensive literature review of voltage control methods in  distribution systems using distributed energy resources and limit the literature review below to the studies most relevant to our work.

Turitsyn et al. \cite{kostya_2011} describe and compare a set of decentralized control policies of distributed inverters for voltage control and power flow loss minimization purposes. The numerical experiments in \cite{kostya_2011} suggests that a hybrid control policy that trade-offs voltage and loss minimization objectives is more effective that either of single-objective control policies. Furthermore, reference \cite{kostya_2011}  demonstrates that the performance of any control policy improves as the number of input measurements (active power, reactive power, voltage magnitude) increases. In \cite{compare} and \cite{Kundu_2013}, the authors integrate the control policies similar to \cite{kostya_2011} in the distribution power flow optimization and compare the centralized and decentralized operating frameworks. Even though the centralized framework yields a slightly less expensive operating cost, it requires ubiquitous communication channels, which is impractical and can be seen as an undesirable vulnerability. On the other hand, the decentralized framework requires no communication and is capable of meeting operating limits on power flows and voltages within their respective ranges at an acceptable cost, \cite{Kundu_2013}. Šulc et al. \cite{Petr_ADMM} attempt to bridge the gap between the centralized and decentralized frameworks by enabling communication between neighboring distributed energy resources with the control policy from \cite{kostya_2011}. Limited communications between the neighboring distributed energy resources improves the cost performance of the decentralized framework at expense of solving an iterative algorithm based on the alternating direction method of multipliers (ADMM) \cite{ADMM}.

This paper  continues the work in \cite{kostya_2011} and \cite{Petr_ADMM}. We extend the control policy from  \cite{kostya_2011} to react to changes in the power flow at the connection point of distributed inverters. Next, we integrate this policy in the distribution power flow optimization with the limited communication as in \cite{Petr_ADMM} and extend the optimization to account for the uncertainty of PV injections using chance constraints. The chance-constrained framework has previously been applied in the context of power flow optimization under uncertainty in both transmission \cite{chance_bienstock,chance_SOC, cc_robust_2017} and distribution  systems \cite{baker_2017}. \textcolor{black}{In this work, the use of chance constraints is motivated by the need to accommodate stochasticity of PV resources in a convex form that is computationally tractable and allows for ADMM-like solution techniques.} The proposed distribution power flow optimization with chance constraints is solved in a decentralized manner using an iterative ADMM-like algorithm inspired by \cite{Petr_ADMM}. Our case study is performed on the IEEE 33-bus distribution test system and compares the centralized and decentralized framework under the proposed power-flow-based control policy.

The remainder of this paper is organized as follows. Section \ref{sec:review} describes existing control policies for distributed inverters and the proposed control policy.  Section \ref{sec:opt} describes how the proposed control policy and the ADMM-like algorithm are integrated in the distribution power flow optimization with chance constraints. Section \ref{sec:conclusion}  concludes the paper.

\section{Decentralized Control Policies for Distributed Inverters} \label{sec:review}
\subsection{Preliminaries}
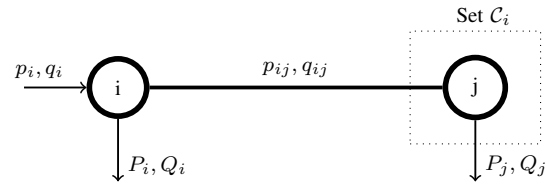
\begin{figure}[!b]
        \captionsetup{labelsep=period, font=footnotesize}
\centering
    \begin{tikzpicture}[auto, node distance=1cm,auto, font=\footnotesize]
    \tikzstyle{block} = [circle, draw, fill=white!15, text width=0.4cm, text centered,  minimum height=0.4cm]
    \tikzstyle{block_rect} = [rectangle, draw, dotted, fill=white!10, text width=1.5cm, text centered,  minimum height=1.5cm]
    
    \node [block, line width=0.75mm] (b1) {i} ;
    \node [block_rect, right of=b1, xshift=3.75cm ] (b3) {} ;
    \node [block, right of=b1, xshift=3.75cm, line width=0.75mm ] (b2) {j} ;
    \node[text width=3cm] at (6.00,0.95) 
    {Set $\mathcal{C}_i$};
    
	\coordinate[below of=b1, yshift=-0.25cm] (c1);
	\coordinate[left of=b1, xshift=-0.25cm] (c2);
	
	\coordinate[below of=b2, yshift=-0.25cm] (c3);
	\coordinate[right of=b2, xshift=0.25cm] (c4);

    \draw [-,thick, line width=0.5mm] (b1) -- (b2) node [midway] {$p_{ij}, q_{ij}$};
    \draw [->,thick, line width=0.25mm] (b1) -- (c1) node [near end] {$P_{i}, Q_{i}$};
    \draw [->,thick, line width=0.25mm] (c2) -- (b1) node [near start] {$p_{i}, q_{i}$};
   \draw [->,thick, line width=0.25mm] (b2) -- (c3) node [near end] {$P_{j}, Q_{j}$};

\end{tikzpicture}
\vspace{3pt}
\caption{A radial distribution system with two nodes and main notations.}
\label{fig_1}
\end{figure}

We consider a radial distribution system represented by graph  $\mathcal{G} = (\mathcal{E}, \mathcal{N})$, where $\mathcal{E}$ and $\mathcal{N}$ are the sets of edges (lines) and nodes (buses), see Fig.~\ref{fig_1}. The set of nodes where PV resources are located is denoted as $ \mathcal{N}^{PV} \in \mathcal{N}$.  Each node is characterized by its active and reactive load ($P_i$ and $Q_i, i \in \mathcal{N}$), its active and reactive PV generation ($p_i$ and $q_i, i \in \mathcal{N}$), and its voltage magnitude $v_i \in [\underline{V}_i, \overline{V}_i],  i \in \mathcal{N}$, where $\underline{V}_i$ and $\overline{V}_i$ are the nodal voltage limits.  It is assumed that PV resources are connected via an inverter with the rated power of $S_i$.  The active and reactive power flows from node $i$ to node $j$ are denoted as $p_{ij}$ and $q_{ij}, (ij) \in \mathcal{E}$, and the resistance and reactance of edge $(ij)$ are $R_{ij}$ and $X_{ij}$. Additionally, set  $\mathcal{C}_i$ denotes all children nodes of node $i$.

Neglecting line losses, power flows in a radial distribution system can be modeled using the \textit{LinDistFlow } approximation, \cite{bw_1989}:
\begin{flalign}
    p_{ij} = P_j - p_j + \sum_{k \in \mathcal{C}_j} p_{jk}, \quad \forall (ij) \in \mathcal{E} \label{flow_p} \\
        q_{ij} = Q_j - q_j + \sum_{k \in \mathcal{C}_j} q_{jk}, \quad \forall (ij) \in \mathcal{E}  \label{flow_q}  \\
        u_j = u_i - 2 (R_{ij} p_{ij} + X_{ij} q_{ij}), \quad \forall (ij) \in \mathcal{E}   \label{volt_sq}  
\end{flalign}
where \eqref{flow_p} and \eqref{flow_q} govern the reactive and active power flows and \eqref{volt_sq} accounts for the nodal voltages squared, i.e.  $u_i=v_i^2$ and $u_j=v_j^2$.

The active and reactive power injection of PV resources are constrained by the rated power of the inverter:
\begin{flalign}
    p_i^2+q_i^2 \leq S_i^2, \quad \forall i \in \mathcal{N}^{PV}. \label{inverter_limit_max}
\end{flalign}
If the active power output is fixed at $\hat{p}_i$, the reactive power out of the inverter can vary as:
\begin{flalign}
    |q_i| \leq \sqrt{S_i^2 - \hat{p}_i^2}, \quad \forall i \in \mathcal{N}^{PV}. \label{inverter}
\end{flalign}
The ability to control the reactive power output of distributed inverters within the range given by \eqref{inverter} enables various control policies for voltage support and loss minimization. These policies are reviewed in Section~\ref{sec:policies}.

\subsection{Existing Control Policies} \label{sec:policies}
This section reviews existing control policies that are grouped below based on their intended purpose.

\subsubsection{Constant power factor}
This policy aims to maintain a constant power factor of net power injections at a given node of the distribution system, i.e. $\cos \phi_i = \text{const}, \forall i \in \mathcal{N} $. This policy is based on local reactive and active power measurements and adjust the output of the inverter to compensate for the power factor fluctuations. Assuming that active power produced by PV resources cannot be curtailed or stored for later use, i.e. $p_i = \hat{p}_i$, the \textcolor{black}{inverter} at node $i$ adjusts its reactive power output as:
\begin{flalign}
 q_i = \big(  (\hat{p}_i-P_i)^2+Q_i^2\big) /\cos \phi_i^2 - (\hat{p}_i-P_i)^2 , \quad \forall i \in \mathcal{N}^{PV}. \label{const_pf_pol}
\end{flalign}
where $q_i$ and $\hat{p}_i$ are subject to \eqref{inverter_limit_max}. Note that \textcolor{black}{in} some cases, the constant power factor policy can be relaxed such that the power factors needs to be maintained within a given range, rather than at a given constant value \cite{pfc_policy}. In practice, the constant power factor policy has been proven to be efficient to  keep nodal voltages within an acceptable range \cite{kostya_2011}. However, the overall performance of this policy significantly reduces as the number of inverters increases.

\subsubsection{Voltage control}
On the other hand, nodal voltages can be explicitly factored in the control policy for distributed inverters \cite{voltage_control}. References \cite{voltage_control,kostya_2011, voltage_control2} study variations of such policies based on a so-called sigmoid function, where the control law is given by:
\begin{flalign}
 q_i = F^V_i=\min \big[ Q_i, \sqrt{S_i^2 - \hat{p}_i^2} \cdot \text{sgn} (v_i, \delta)  \big] , \quad \forall i \in \mathcal{N}^{PV} \label{const_voltage}
\end{flalign}
where $\text{sgn} (\cdot)$ denotes the sigmoid  operator and $\delta$ is an exogenous parameter that can be tuned in each specific case. Relative to the control policy in \eqref{const_pf_pol}, the control policy in \eqref{const_voltage} is more effective as it directly targets nodal voltages. However, it tends to increase power losses in some cases and, thus, is commonly perceived as less cost effective, see  \cite{kostya_2011}.

\subsubsection{Power loss minimization}
Turitsyn et al. \cite{loss_min_policy} also propose the control policy to minimize power losses: 
\begin{flalign}
 q_i = F^L_i= \begin{cases} Q_i,  \quad |q_i| \leq \sqrt{S_i^2 - \hat{p}_i^2}  \\ \pm \sqrt{S_i^2 - \hat{p}_i^2}, \text{ otherwise} \end{cases} , \quad \forall i \in \mathcal{N}^{PV}. \label{const_losses}
\end{flalign}
In contrast with the policy in \eqref{const_voltage}, the power loss minimization policy may impair the compliance with voltage limits. 

\subsubsection{Hybrid policy} The drawbacks of the control policies in \eqref{const_voltage} and \eqref{const_losses} can be overcome, if the two policies are combined in a hybrid policy, i.e.  $H_i= K^L_i F^L_i + K^V_i F^V_i $, where $K^V_i$ and $K^L_i$ are appropriately chosen droop parameters.  As shown in \cite{kostya_2011}, the hybrid policy outperforms \textcolor{black}{$F^L_i$}  and $F^V_i$, when they are enforced individually.

\subsection{Proposed Control Policy} \label{sec:proposed_policy}
The common thread of the control policies reviewed in Section~\ref{sec:policies} is that they are based on local voltage and power injection measurements, i.e. the response of these policies is limited to the operating conditions at node $i$. In addition to these measurements,  it is also possible to measure downstream active and reactive power flows from node $i$ to its children nodes. Since the distribution system is assumed to be radial, the downstream power flow measurements can be to respond to operating conditions downstream of the node where the measurements are taken. For example, if $q_{ij}$ increases, it indicates that more reactive power is needed downstream of node $i$ and that the inverter at node $i$ can increase its reactive power output $q_{i}$ to compensate for the increase in $q_{ij}$. As a result, less reactive power would be sourced from the upstream of node $i$, naturally leading to lower power flows and voltage sags across the upstream edges. This observation suggests that under the fixed active power injection of the inverter, i.e. $p_i = \hat{p}_i$,  local measurements of $q_{ij}$ can be used to design the following control policy:

\begin{flalign}
 &   q_i = \begin{cases}
\hat{q}_{i} + K^{q}_i \sum_{j \in \mathcal{C}_i} (q_{ij} - \hat{q}_{ij}),  &\text{ if } \hat{q}_{ij} - q_{ij} \geq 0\\
\hat{q}_{i}, &\text{ if } \hat{q}_{ij} - q_{ij} <0
\end{cases}  \label{inverter2}, \\
&    |q_i| \leq \sqrt{S_i^2 - \hat{p}_i^2}, \quad \forall i \in \mathcal{N}^{PV} \label{inverter3}
\end{flalign}
where $K^{q}_i >0 $ is a droop coefficient of the inverter at node $i$, $\hat{q}_{i}$ and $\hat{q}_{ij}$ are the reference reactive power output of the \textcolor{black}{inverter} at node $i$ and power flow in edge $(ij)$. The control enabled by \eqref{inverter2} proportionally increases the reactive power output of inverter at node $i$, $q_{ij}$,  if the measured reactive power flow in edge $(ij)$, $q_{ij}$, is greater than its  reference value $\hat{q}_{ij}$, i.e. $q_{ij}\geq\hat{q}_{ij}$. On the other hand, if $q_{ij}<\hat{q}_{ij}$, the control in \eqref{inverter2} maintains the reference reactive power output $\hat{q}_i$. The feasibility of the control policy is ensured in \eqref{inverter3}.

\subsubsection*{Remark 1}  The control in \eqref{inverter2} is based on reactive power flow measurements and is also  extensible to active power measurements. In this case, the active and reactive power outputs of the inverter are governed by:
\begin{flalign}
&  \text{Eq. } \eqref{inverter2} \label{inverter4} \\
 &   p_i = \begin{cases}
\hat{p}_{i} + K^p_i \sum_{j \in \mathcal{C}_i} (p_{ij} - \hat{p}_{ij}),  &\text{ if } \hat{p}_{ij} - p_{ij} \geq 0\\
\hat{p}_{i}, &\text{ if } \hat{p}_{ij} - p_{ij} <0
\end{cases}  \label{inverter5}, \\
&    p_i^2 + q_i^2 \leq S_i^2 , \quad \forall i \in \mathcal{N}^{PV}, \label{inverter6}
\end{flalign}
where $K^{p}_i >0 $ is a droop coefficient and $\hat{p}_{ij}$ is a reference power output of the inverter. Eq.~\eqref{inverter5} changes the active power output of the inverter at node $i$ similarly to \eqref{inverter2}. Eq.~\eqref{inverter6} enforces the limit on the feasible range of $p_i$ and $q_i$. Since the active power output of PV resources is uncertain and there is no guarantee that it can be increased on demand, enabling the control policy in \eqref{inverter5} will require a coordination with storage resources at the same node $i$.

\subsubsection*{Remark 2}Unlike the control policies reviewed in Section~\ref{sec:policies}, the control in \eqref{inverter2} and in \eqref{inverter5}  require reference values of $\hat{p}_i$, $\hat{q}_i$, $\hat{p}_{ij}$,  $\hat{q}_{ij}$. In case if there is no communication infrastructure between the distribution system operator and distributed inverters, these reference values can be set to some ad-hoc values that can be related to power flow limits for $\hat{q}_{ij}$ or representative power outputs of the inverter for $\hat{p}_i$ and $\hat{q}_i$. On the other hand, if there is a communication mechanism between the system operator and distributed inverters, the reference values can be routinely updated to follow the operating settings of the distribution system operator. If these settings are obtained from solving the  distribution  optimal power flow problem, the controls in \eqref{inverter2} and in \eqref{inverter5} can be interpreted as the minimizers of real-time deviations from the optimal power flow solution.  

\subsubsection*{Remark 3} As with the reference values discussed in Remark 2, droops $K^{p}_i $ and $K^{q}_i$ can be set to fixed parameters that account for average operating conditions at node $i$ or be routinely updated by the distribution system operator, if there is necessary communication infrastructure. 

\subsection{Numerical Experiments}
\begin{figure}[t!]
    \centering
        \captionsetup{labelsep=period, font=footnotesize}
\begin{forest}for tree={circle,draw,minimum size=1.5em, 
            inner sep=1pt,l sep=0.1cm}
 [1,before computing xy={l=15mm,s=0mm}
 [2
    [3,before computing xy={l=10mm,s=0mm}
        [4,before computing xy={l=10mm,s=0mm}
        [5,fill=yellow
        [6,
            [7,before computing xy={l=0mm,s=-10mm}
            [8,before computing xy={l=-10mm,s=0mm}
            [9,before computing xy={l=-10mm,s=0mm}
            [10,before computing xy={l=0mm,s=-10mm}
            [11,before computing xy={l=0mm,s=-10mm}
            [12
            [13
            [14
            [15,before computing xy={l=0mm,s=10mm}
            [16,before computing xy={l=0mm,s=10mm}
            [17,before computing xy={l=0mm,s=10mm}
            [18,before computing xy={l=0mm,s=10mm}]]]]]]]]]]]]
            [26,before computing xy={l=0mm,s=10mm}
            [27,before computing xy={l=-10mm,s=0mm}
            [28,before computing xy={l=-10mm,s=0mm}
            [29,before computing xy={l=-10mm,s=0mm}
            [30,before computing xy={l=0mm,s=10mm}
            [31,before computing xy={l=0mm,s=10mm}
            [32,before computing xy={l=0mm,s=10mm}
            [33,fill=yellow,before computing xy={l=10mm,s=0mm}]]]]]]]]]]]
        [23,before computing xy={l=0mm,s=-10mm}
            [24,before computing xy={l=0mm,s=-10mm}
            [25,before computing xy={l=0mm,s=-10mm}]]]]
    [19,before computing xy={l=0mm,s=10mm}
    [20,before computing xy={l=0mm,s=10mm}
    [21,before computing xy={l=0mm,s=10mm}
    [22,before computing xy={l=0mm,s=10mm}]]]]]]
\end{forest}
\vspace{7pt}
\caption{The 33-bus distribution system described in \cite{bw_1989} and available in Matpower as `case33bw' \cite{matpower}. Node 5 features one inverter-tied PV resource. Cases I and II consider a 50\% load increase at nodes 5 and 33 (both nodes are in yellow), respectively. }
\label{network_diagram}
  \end{figure}
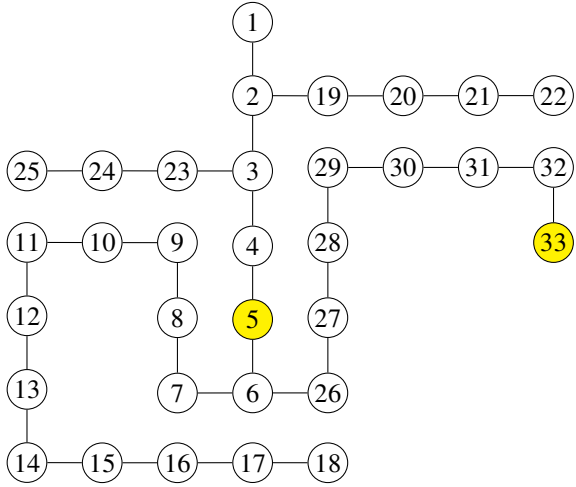

The  policies \eqref{inverter2}-\eqref{inverter3} and \eqref{inverter4}-\eqref{inverter5} are tested using a modification of the IEEE 33-bus distribution system  \cite{bw_1989} available in Matpower as `case33bw' \cite{matpower}, see Fig.~\ref{network_diagram}.  The input data from \cite{matpower} is modified as follows.  One PV generator is connected to the distribution grid at node 5 via the inverter with the rated  power $S = 0.5$ MVA. The power output of the PV generator is fixed at the reference value of $\hat{p}_i=0.3$ MW, thus $|q_i| \leq 0.4$ MVAr. 
The PV resource  is coupled with one storage unit that can provide up to 0.1 MW. The reference values of $\hat{q}_{ij}$ and $\hat{p}_{ij}$ are obtained from the AC OPF solution obtained with Matpower. For the sake of simplicity, it is assumed that $K=K_i^{p}=K_i^{q}$.  The active and reactive power loads are fixed at all nodes, except for two cases. In Case I the load at node $5$ is increased by 50\%, while in Case II the load at node $33$ (the most electrically remote node from the root bus of the distribution system) is increased by 50\%. The motivation behind these two cases is to compare how different control policies perform under local and remote deviations.

\begin{figure}[t!]
        \captionsetup{labelsep=period, font=footnotesize}
  \centering
\includegraphics[width=\columnwidth]{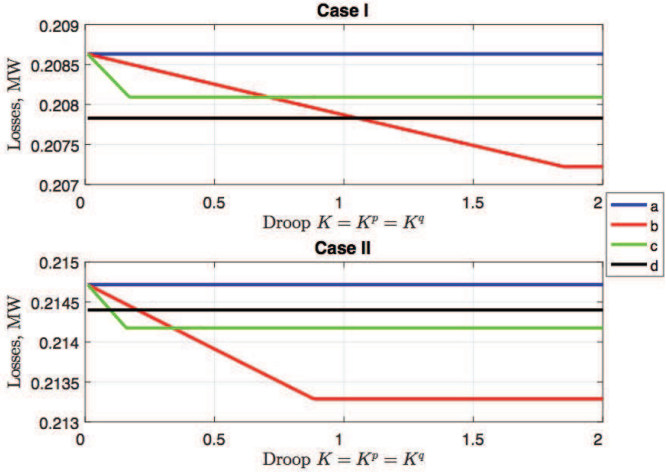}
\vspace{1pt}
    \caption{Comparison of the total active power losses in the distribution system: a) no control,  \textcolor{black}{b) control policy in~\eqref{inverter4}-\eqref{inverter6}, c) \textcolor{black}{control} policy in~\eqref{inverter2}-\eqref{inverter3}} , d) $F_i^{L}$ in \eqref{const_losses}. }
        \label{fig:comparison_of_policies}
     \end{figure}

Fig.~\ref{fig:comparison_of_policies} shows how the proposed policy changes the total active power losses in the distribution system for different values of droop $K$ and compares it to the case without any control from the inverter and with the power loss minimization policy in \eqref{const_losses}.  As expected, using any control policy reduces the losses relative to the case when no inverter-based control is enabled. In both cases, the comparative performance of the proposed policies relative to \eqref{const_losses} is sensitive to the value of droop $K$ and the proposed policies become more advantageous as the value of $K$ increases. With respect to the value of droop $K$, the proposed policies linearly reduce the losses for higher values of $K$ until the output of the inverter achieves $S$, leading to a breakpoint. Following the breakpoint,  increasing the value of $K$ does not result in any additional power loss reduction. The control policy in Eq.~\eqref{inverter4}-\eqref{inverter6}, which regulates both the active and reactive power output of the inverter, saturates at higher values of droop $K$ than the control policy in \eqref{inverter2}-\eqref{inverter3} and thus leads to larger power loss reductions.

\section{Control Policies in Chance-Constrained Decentralized Optimization} \label{sec:opt}
\textcolor{black}{This} section describes how the proposed control policies in  \eqref{inverter2}-\eqref{inverter3} and \eqref{inverter4}-\eqref{inverter5} can be enforced in the distribution optimal power flow problem.  The resulting problem is then solved using the ADMM-like algorithm inspired by \cite{Petr_ADMM} in a distribution fashion that mirrors the distributed nature of the proposed control policies. Finally, we extend the problem to accommodate the uncertainty of PV resources using the chance-\textcolor{black}{constrained} framework.

\subsection{Formulation}
The distribution optimal power flow problem is formulated as:
\begin{align} 				
&\underset{p,q}{\text{min}} \sum_{(ij) \in \mathcal{E}} R_{ij} \frac{(p_{ij})^2+(q_{ij})^2}{V_i^2} \label{ldf1}\\
&\text{Eq. }\eqref{flow_p}-\eqref{inverter} \label{ldf22}
\end{align}
where the objective function is to minimize the power losses and $V_i$ is the nominal voltage at node $i$. \textcolor{black}{Since \textit{LinDistFlow} neglects second-order terms of power flow equations, \eqref{ldf1} computes approximate power losses based on approximate values of power flows  $p_{ij}$ and $q_{ij}$.} To solve this problem in a decentralized manner,  we reformulate it using the ADMM consensus approach as elaborated in \cite{Petr_ADMM}:  
\begin{align} 				
&\underset{q}{\text{min}} \sum_{(ij) \in \mathcal{E}} \frac{R_{ij}(q_i^-)^2}{V_i^2} \label{admm1}\\
& q_i^+ - q_i^- + Q_i - |q_i| \leq 0,  \quad \forall i \in \mathcal{N} \label{admm2} \\
& -q_i^+ + q_i^- - Q_i - |q_i| \leq 0, \quad \forall i \in \mathcal{N}  \label{admm3} 
\end{align}
\begin{align} 
& \underline{V}_i^2  \leq u_i^+ \leq \overline{V}_i, \forall i \in \mathcal{N}   \label{admm5}  \\
& u_{i}^+ = u_i^- - 2(R_{ij}p_{ij} + X_{ij}q_{i}^-), \quad \forall (i,j) \in \mathcal{E}  \ \label{admm6} \\
& u_{i}^+ = u_i, \ u_{i}^- = u_{j}, \quad \forall (i,j) \in \mathcal{E}  \label{admm7} \\
& q_i^+ = q_i, \ q_i^- = q_{j},  \quad \forall i \in \mathcal{N}  \label{admm4} \\
& \text{Eq.}~\eqref{inverter4}-\eqref{inverter6} \label{admm8} 
\end{align}
where $q_i^+, q_i^-, u_i^+$ and $u_i^-$ are the local copies of the global variables $q_i, q_{ij}, u_i$ and \textcolor{black}{$u_{j}$} for each node $j$, i.e. \textcolor{black}{only communication between neighboring nodes is required.} Eqs.~\eqref{admm7} and \eqref{admm4} represent consensus or coupling constraints ensuring that all the local copies of the variables are equal to the global variables. Eq.~\eqref{admm8} enforces the proposed control policies.

\subsection{Solution Technique}
Following \cite{Petr_ADMM}, we implement the consensus ADMM-like algorithm by relaxing the consensus constraints \eqref{admm4} and \eqref{admm7}. Each iteration of the algorithm is detailed below:
\begin{enumerate}

\item \text{Minimization step}:
This step solves the local optimization problem for every PV node, where the objective function minimizes the augmented Lagrangian of the consensus problem: 
\begin{align}
\begin{split}
& \underset{q_i^-,q_i^+,u_i^-,u_i^+}{\text{min}}\sum_{(ij) \in \mathcal{E}} \frac{R_{ij}(q_i^-)^2}{V_i^2} \\& + \frac{\rho}{2}(q_i^+ - q_i)^2 + \frac{\rho}{2}(q_i^- - q_{ij})^2 \\& + \frac{\rho}{2}(u_i^+ - u_i)^2 + \frac{\rho}{2}(u_i^- - u_{j})^2 \\& + \lambda_i^{q^+}(q_i^+ - q_{ij}) + \lambda_i^{q^-}(q_i^- - q_{ij}) \\& + \lambda_i^{u^+}(u_i^+ - u_i) + \lambda_i^{u^-}(u_i^- - u_{j}) \label{min1}
\end{split} \\
&\text{Eq. }\eqref{admm2}-\eqref{admm6}, \eqref{admm8} \label{min2}
\end{align}
where $\frac{\rho}{2}$ represent penalties for the difference between the local and global variables and $\lambda_i^{q^+}, \lambda_i^{q^-}, \lambda_i^{u^+}$ and $\lambda_i^{u^-}$ are the dual variables associated with  \eqref{admm7} and \eqref{admm4}. The minimization step yield $q_i^+(k+1), q_{i+1}^-(k+1), [u_i^+(k+1), u_{i+1}^-(k+1)$ at iteration $k$. 

\item \text{Global variables update step}:
Communicating between the neighbouring nodes only, the global variables are updated as:
\begin{align}
& q_i (k+1) = \frac{1}{2}\big[q_i^+(k+1) + q_{j}^-(k+1)\big] \label{avg1} \\
& u_i (k+1) = \frac{1}{2}\big[u_i^+(k+1) + u_{j}^-(k+1)\big]. \label{avg2} 
\end{align}

\item \text{Update step}:
This step updates the dual variables for every node as:
\begin{align}
& \lambda_i^{q^+}\!\!(k+1) \!=\! \lambda_i^{q^+}\!\!(k)\! +\! \rho\big[q_i^+\!(k+1) \!-\! q_i\!(k+1)\big] \label{update1} \\ 
& \lambda_i^{q^-}\!\!(k+1) \!=\! \lambda_i^{q^-}\!\!(k) \!+\! \rho\big[q_i^-\!(k+1) \!-\! q_{ij}\!(k+1)\big] \label{update2} \\
& \lambda_i^{u^+}\!\!(k+1) \!=\! \lambda_i^{u^+}\!\!(k) \!+\! \rho\big[u_i^+\!(k+1)\! -\! u_i\!(k+1)\big] \label{update3} \\
& \lambda_i^{u^-}\!\!(k+1) \!= \!\lambda_i^{u^-}\!\!(k)\! +\! \rho\big[u_i^-\!(k+1) \!- \!u_{ij}\!(k+1)\big]. \label{update4}
\end{align}
\end{enumerate}
Finally, the net reactive power injected or consumed by the inverters for each node is recovered from the obtained solution:
\begin{flalign}
 q_i = q_i^+ - q_i^- + Q_i. \label{reactive_injection} 
\end{flalign}

\subsection{Modeling PV Uncertainty}
The PV uncertainty can be considered in the proposed distribution optimal power flow problem using the chance-constrained framework. Following the chance-constrained approach in \cite{chance_bienstock,chance_SOC},  we obtain:
\begin{align}
& \mathbb{P}({p}_i \leq \overline{P}_i) \geq 1 - \epsilon, \quad \forall i \in \mathcal{N}^{PV} \label{cc1} \\
& \mathbb{P}({p}_i \geq \underline{P}_i) \geq 1 - \epsilon, \quad \forall i \in \mathcal{N}^{PV} \label{cc2}
\end{align}
where $\big[ \underline{P}_i, \overline{P}_i \big]$ is a given forecast range and  $\epsilon$ is a given tolerance to a violation of that range. Accordingly, \eqref{admm2} and \eqref{admm3} are modified as:
\begin{align}
& \mathbb{P}(q_i^+ - q_i^- + Q_i - \cos \phi_i p_i \leq 0) \geq 1 - \epsilon, \quad \forall i \in \mathcal{N}^{PV} \label{pf1} \\
& \mathbb{P}(-q_i^+ + q_i^- - Q_i - \cos \phi_i p_i \leq 0) \geq 1 - \epsilon, \quad \forall i \in \mathcal{N}^{PV}  \label{pf2}
\end{align}
where $\cos \phi_i$ represents a constant power factor. We can also introduce the chance constraints on voltage limits  in  \eqref{admm5}:
\begin{align}
& \mathbb{P}( u_i^+ \leq \overline{V}_i )  \geq 1 - \epsilon, \quad \forall i \in \mathcal{N}^{PV} \label{vcc1} \\
& \mathbb{P}( u_i^+ \geq \underline{V}_i  \geq 1 - \epsilon, \quad \forall i \in \mathcal{N}^{PV} \label{vcc2}
\end{align}

Note that chance constraints in \eqref{cc1}-\eqref{vcc2} can be reformulated as second order cone (SOC) constraints as explained in \cite{chance_bienstock,chance_SOC}. However, in our implementation we treat \eqref{vcc1}-\eqref{vcc2} deterministically to avoid complexity.  Interested readers are referred to Appendix for our description of this transformation.  \textcolor{black}{Distributed implementation of \eqref{vcc1}-\eqref{vcc2} will be a focus of our future work. }

\subsection{Numerical Experiments}
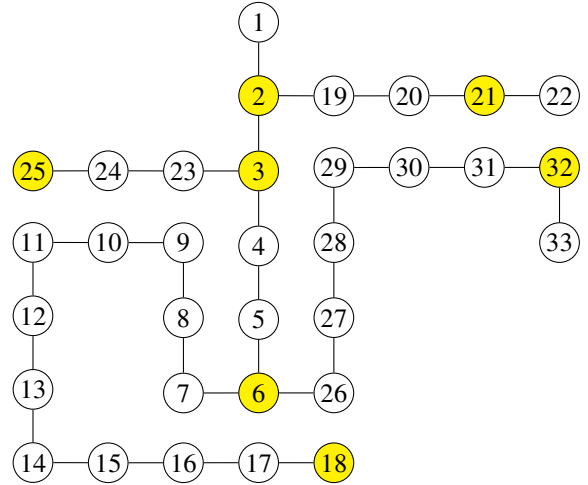
\begin{figure}[t!]
    \centering
    
            \captionsetup{labelsep=period, font=footnotesize}
\begin{forest}for tree={circle,draw,minimum size=1.5em, 
            inner sep=1pt,l sep=0.1cm}
 [1,before computing xy={l=15mm,s=0mm}
 [2,fill=yellow,
    [3,fill=yellow,before computing xy={l=10mm,s=0mm},
        [4,before computing xy={l=10mm,s=0mm}
        [5
        [6,fill=yellow,
            [7,before computing xy={l=0mm,s=-10mm}
            [8,before computing xy={l=-10mm,s=0mm}
            [9,before computing xy={l=-10mm,s=0mm}
            [10,before computing xy={l=0mm,s=-10mm}
            [11,before computing xy={l=0mm,s=-10mm}
            [12
            [13
            [14
            [15,before computing xy={l=0mm,s=10mm}
            [16,before computing xy={l=0mm,s=10mm}
            [17,before computing xy={l=0mm,s=10mm}
            [18,fill=yellow,before computing xy={l=0mm,s=10mm}]]]]]]]]]]]]
            [26,before computing xy={l=0mm,s=10mm}
            [27,before computing xy={l=-10mm,s=0mm}
            [28,before computing xy={l=-10mm,s=0mm}
            [29,before computing xy={l=-10mm,s=0mm}
            [30,before computing xy={l=0mm,s=10mm}
            [31,before computing xy={l=0mm,s=10mm}
            [32,fill=yellow,before computing xy={l=0mm,s=10mm}
            [33,before computing xy={l=10mm,s=0mm}]]]]]]]]]]]
        [23,before computing xy={l=0mm,s=-10mm}
            [24,before computing xy={l=0mm,s=-10mm}
            [25,fill=yellow,before computing xy={l=0mm,s=-10mm}]]]]
    [19,before computing xy={l=0mm,s=10mm}
    [20,before computing xy={l=0mm,s=10mm}
    [21,fill=yellow,before computing xy={l=0mm,s=10mm}
    [22,before computing xy={l=0mm,s=10mm}]]]]]]
\end{forest}
\vspace{5pt}
\caption{The modified 33-bus distribution system where the nodes with the PV resources installed at nodes \#2,3,6,18,21,25 and 32 (in yellow).
}
\label{network_diagram2}
  \end{figure}

We evaluate the proposed decentralized chance-constrained distribution optimal power optimization on the IEEE  33-bus distribution system, where the PV resources are installed at nodes \#2 (1.9 MW), 3 (3.77MW),6 (7.54 MW), 18 (1.88 MW), 21 (4.71 MW), 25 (4.24 MW) and 32 (5.94 MW), see Fig.~\ref{network_diagram2}.  The forecast error at every PV resources is zero-mean with the forecast varies of 10\% of the forecast output. We assume that parameter $\rho = 1/V_1^2$, where $V_1$ is the nominal voltage at the root bus. In the following experiments, the value of $\epsilon$ is uniformly set for all chance constraints. The proposed decentralized optimization is compared to the deterministic centralized (global) optimization. 

\subsubsection{Power Losses}

\begin{figure}[!t]
\centering        \captionsetup{labelsep=period, font=footnotesize}
\includegraphics[trim={0 0 0 0},width=\columnwidth]{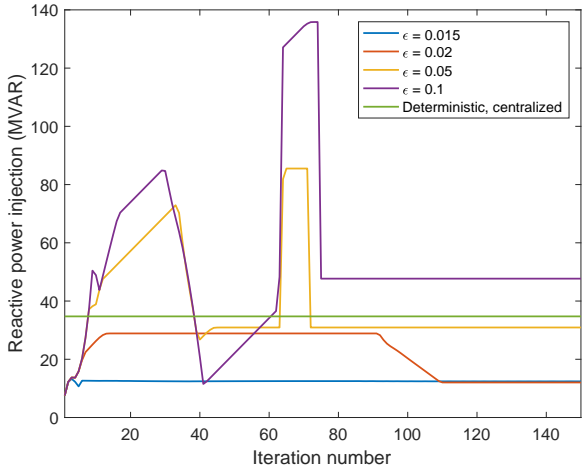}
\vspace{5pt}
\caption{Comparison of the \textcolor{black}{ADMM} algorithm convergence in terms of the total reactive power injection of distributed PV inverters  for different values of  $\epsilon$. The straight green line indicates  the deterministic centralized case. }

\label{react_all}
\end{figure}

\begin{figure}[t]
\centering
\includegraphics[trim={0 0 0 0},width=\columnwidth]{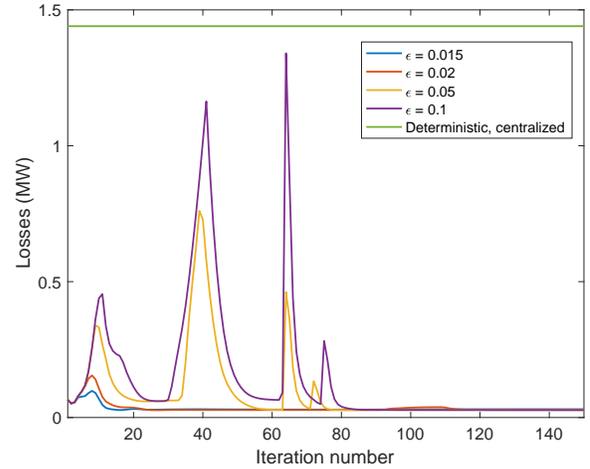}
        \captionsetup{labelsep=period, font=footnotesize}
        \vspace{5pt}
\caption{Comparison of the \textcolor{black}{ADMM} algorithm convergence in terms of the system-wide active power losses for different values of  $\epsilon$. The straight green line indicates  the deterministic centralized case. }
\label{losses}
\end{figure}

Fig.~\ref{react_all} and \ref{losses} display the ADMM convergence in terms of the total reactive power injection of distributed inverters and system-wide active power losses. Even though the algorithm takes no more than 120 iterations to converge among all instances considered, there are distinguished spikes in both plots. These spikes are characterized by the breakpoints, similar to Fig~\ref{fig:comparison_of_policies}, when the  inverter is operated at its rated capacity and can no longer provide the reactive power support to the system. As the value of parameter $\epsilon$ increases, i.e. the chance constraints become tighter, the total reactive power injection of distributed  inverters increases monotonically. This indicates that the proposed control policy and decentralized optimization are of greater value for stricter operating standards. On the other hand, the system-wide active power losses remains roughly the same for all chance constrained instances solved. The effect of the proposed decentralized optimization is particularly noticeable for the system-wide active power losses, which drastically reduce relatively to the deterministic centralized case, regardless of the $\epsilon$ value chosen.

\subsubsection{Nodal voltages}

\begin{figure}[!t]
\centering
\includegraphics[trim={0 0 0 0},width=\columnwidth]{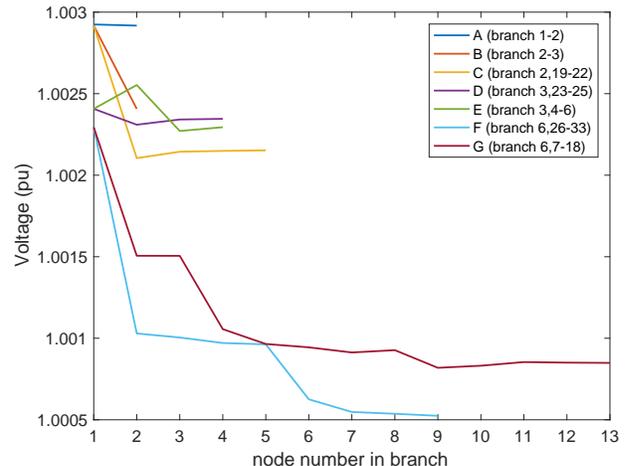}
        \captionsetup{labelsep=period, font=footnotesize}
\caption{Voltage profile along different branches of the  distribution system for the decentralized optimization,  parameter $\epsilon$ = 0.05. The branches are numbered based on Fig.~\ref{network_diagram2}.}
\label{voltage_leaves}
\end{figure}

\begin{figure}[!t]
\centering
        \captionsetup{labelsep=period, font=footnotesize}
\includegraphics[trim={0 0 0 0},width=\columnwidth]{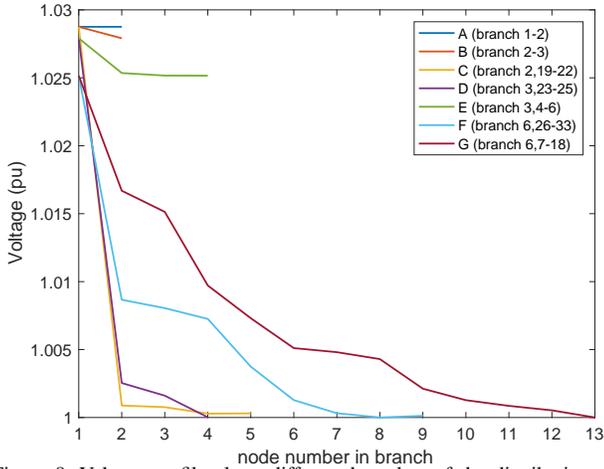}
\caption{Voltage profile along different branches of the  distribution system for the centralized optimization.  The branches are numbered based on Fig.~\ref{network_diagram2}.}
\label{voltage_leaves_glob}
\end{figure}

\begin{figure}[!t]       
\centering
\includegraphics[trim={0 0 0 0},width=\columnwidth]{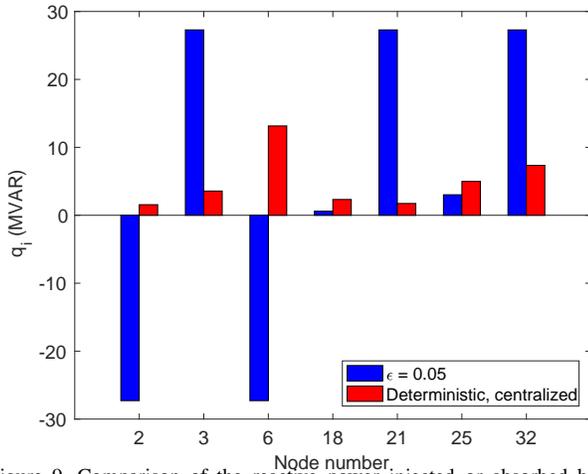}
        \captionsetup{labelsep=period, font=footnotesize}
\caption{Comparison of the reactive power injected or absorbed by every inverter for the centralized problem and the decentralized problem with the value of $\epsilon$ set as 0.05}
\label{react_each}
\end{figure}

This section studies the impact of the proposed decentralized optimization on the nodal voltages across the distribution system. We set the value of parameter $\epsilon$ to 0.05 and track the voltage profile across branches as illustrated in Fig.~\ref{network_diagram2}. Fig.~\ref{voltage_leaves} and \ref{voltage_leaves_glob} detail the voltage profiles for each branch for the decentralized and centralized cases, respectively, while Fig.~\ref{react_each} itemizes the nodal reactive power injections of distributed inverters for both cases. In the centralized case, the voltage profiles monotonically reduce in all branches. On the other hand, the effect of the decentralized optimization is two-fold. First, it reduces the gap between the voltage magnitudes at the starting and end points of each branch, as compared to the centralized case. Second, the voltage profile across some branches is not monotonic. As it can be seen in Fig.~\ref{voltage_leaves} branches C, E, G, F has voltage spikes that are caused by the reactive power injections of distributed  inverters. These spikes are caused by different reactive power injections in the decentralized and centralized cases as in Fig.~\ref{react_each}.

\section{Conclusion}\label{sec:conclusion} 
This paper proposes a power-flow-based control policy for distributed energy resources located in the distribution systems. Our numerical experiments demonstrate that the proposed policy outperforms existing policies in terms of active power loss minimization. The proposed policy is integrated with the decentralized, chance-constrained distribution optimal power flow optimization. As compared to the centralized, deterministic optimization, \textcolor{black}{our method reduces voltage sags across nearly all edges, thus improving compliance with voltage regulation, and reduces power flow losses}. Modeling the PV uncertainty using chance constraints makes it possible to adjust the conservatism of our method and does not increase its computational complexity. The proposed method is extensible to accommodate other control policies and various distribution energy resources.

\appendix

Let $\xi$$\sim$$N(\mu,\Sigma)$ be the vector of random variable with the means and variances given by the vectors $\mu$ and variances $\sigma$, respectively, and let $b$ and $x$ be the vectors of parameters and decision variables, respectively. The chance constraint of the form:
\begin{flalign}
   \mathbb{P}(\xi^T x \leq b) \geq 1 - \epsilon \label{ccgen1} 
\end{flalign}
can be represented in the following form:
 \begin{flalign}
   \mu^Tx + \Phi^{-1}(1-\epsilon) \sqrt{x^T \Sigma x} \leq b    \label{ccgen2} 
\end{flalign}
where $\epsilon \in [0,1]$ is tolerance to violations and $\Phi^{-1}$ is the inverse cumulative distribution function of the standard normal distribution. Eq.~\eqref{ccgen2} can the equivalently be replaced by the following SOC constraints:
\begin{align}
& t \geq \left\Vert {\Sigma^{\frac{1}{2}} x} \right\Vert_2   \label{ccgen3} \\
& \mu^Tx + \Phi^{-1}(1-\epsilon)t \leq b  \label{ccgen4}
\end{align}
In our implementation, this transformation was automatically performed by JumpChance, a Julia package, see \cite{jumpchance}.

\end{document}